\documentclass[prd,preprint,tightenlines,floatfix,showpacs,preprintnumbers,nofootinbib,eqsecnum]{revtex4}

 \usepackage[dvips,final]{graphicx}
  \usepackage{amssymb}
   \usepackage{amsmath}
    \usepackage{amsfonts}
     \usepackage{epsfig}
      \usepackage{bm}

\begin{document}
\thispagestyle{empty} \preprint{\hbox{}} \vspace*{-10mm}

\title{Exclusive double-diffractive production
of open charm \\ in proton-proton and proton-antiproton collisions}

\author{R.~Maciu{\l}a}
\email{rafal.maciula@ifj.edu.pl} \affiliation{Institute of Nuclear
Physics PAN, PL-31-342 Cracow,
Poland}

\author{R.~S.~Pasechnik}
\email{roman.pasechnik@fysast.uu.se} \affiliation{High Energy
Physics, Department of Physics and Astronomy, Uppsala University Box
535, SE-75121 Uppsala, Sweden}

\author{A.~Szczurek}
\email{antoni.szczurek@ifj.edu.pl} \affiliation{Institute of Nuclear
Physics PAN, PL-31-342 Cracow,
Poland and\\
University of Rzesz\'ow, PL-35-959 Rzesz\'ow, Poland}

\date{\today}

\begin{abstract}
We calculate differential cross sections for exclusive double
diffractive (EDD) production of open charm in proton-proton and
proton-antiproton collisions. Sizeable cross sections are found. The
EDD contribution constitutes about 1 \% of the total inclusive cross
section for open charm production. A few differential distributions
are shown and discussed. The EDD contribution falls faster both with
transverse momentum of the $c$ quark/antiquark and the $c \bar c$
invariant mass than in the inclusive case.
\end{abstract}

\pacs{13.87.Ce,14.65.Dw}

\maketitle

\section{Introduction}

The open charm production is often considered as a flag reaction to
test the gluon distributions in the nucleon. For the $c \bar c$ and
$b \bar b$ production at high-energies the gluon-gluon fusion is
assumed to be the dominant mechanism. This process was calculated in
the NLO collinear \cite{NLO_ccbar} as well as in the
$k_t$-factorisation \cite{kt-factorisation_ccbar,BS00,HKSST-qq,LS06} 
approaches by several authors. These analyses seem to report on missing
strength\footnote{The situation is often somewhat clouded by
studying the uncertainty bands due to variation of renormalization
and factorization scales. These analyses lead to rather broad
uncertainty bands which prevent definite conclusions.}. This
suggests that other processes ignored so far should be carefully
evaluated.

\begin{figure}[h!]
\centerline{\epsfig{file=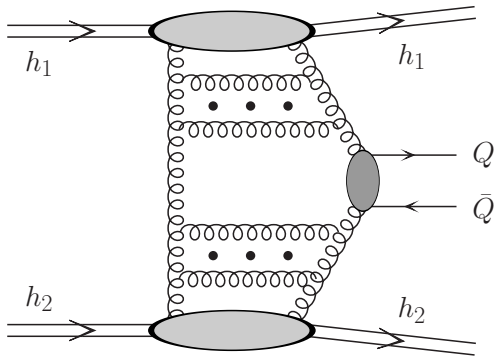,width=7cm}}
\caption{\small The mechanism of exclusive double-diffractive
production of open charm.}
\label{fig:EDD_mechanism}
\end{figure}

The number of potential contributions is not small. In the present
paper we concentrate on exclusive double diffractive (EDD)
mechanism, which was not considered so far for the $c \bar c$
production. The mechanism of the exclusive double-diffractive
production of open charm is shown in Fig.~\ref{fig:EDD_mechanism}.

The EDD $b \bar b$ reaction consitutes a irreducible background to
the exclusive Higgs boson production
\cite{KMR_Higgs_bbbar_background} measured in the $b \bar b$
channel. Up to now only approximate estimates of the $b \bar b$
production were presented in the literature. In the present paper we
consider the $p p \to p p c \bar c$ reaction as a genuine 4-body
process with exact kinematics which can be easily used with
kinematical cuts. The amplitude of the genuine four-body reaction is
written in analogy to the Kaidalov-Khoze-Martin-Ryskin (KKMR)
approach used previously for the exclusive Higgs boson production
\cite{KMR_Higgs,KKMR,KKMR-spin}.

\section{Matrix element and the cross section \\
for exclusive double diffractive $q\bar{q}$ pair production}

Inclusive heavy quark/antiquark pair production was considered in
detail, e.g. in Refs.~\cite{HKSST-qq,LS06}. The nonrelativistic QCD
methods were successfully applied also in the case of central
exclusive production of heavy quarkonia in
Refs.~\cite{PST_chic01,KMR_chic,PST_chic2}. It looks quite natural to apply
similar ideas to exclusive diffractive $q\bar{q}$ (unbound) pair
production.

\subsection{Kinematics}

\begin{figure}[!h]
\centerline{\epsfig{file=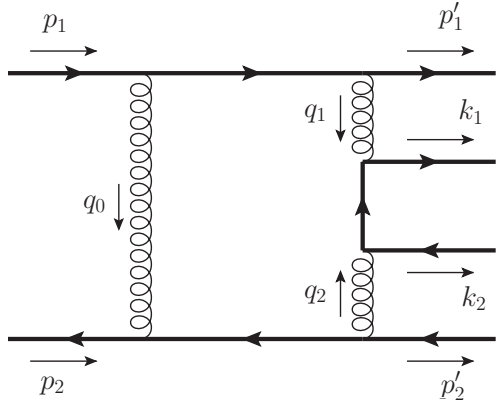,width=7cm}}
\caption{\small
Kinematical variables of exclusive diffractive production
of $q\bar{q}$ pair.}
\label{fig:kinematics_qcd}
\end{figure}

The kinematical variables for the process $pp\rightarrow
p+\mathrm{``gap"}+(q\bar{q})+\mathrm{``gap"}+p$ are shown in
Fig.~\ref{fig:kinematics_qcd}.
%
%

We adopt here the following standard definition of the light cone
coordinates
\[
k^{+}\equiv n^+_{\alpha}k^{\alpha}=k^{0}+k^{3},\text{ }k^{-}\equiv
n^-_{\alpha}k^{\alpha}=k^{0}-k^{3},\text{ }k_{t
}=(0,k^{1},k^{2},0)=(0,{\bf k,}0{\bf )},
\]
where $n^{\pm}$ are the light-cone basis vectors. In the c.m.s.
frame
\begin{eqnarray}
\label{lcbasis} n^+=\frac{p_2}{E_{cms}},\qquad
n^-=\frac{p_1}{E_{cms}},
\end{eqnarray}
and the momenta of the scattering hadrons are given by
\[
p_{1}^{+}=p_{2}^{-}=\sqrt{s},\text{ \ }p_{1}^{-}=p_{2}^{+}=p_{1,t
}=p_{2,t }=0,
\]
with the Mandelstam variable $s=4E_{cms}^2.$

Within the standard $k_t$-factorisation approach, the decomposition
of gluon momenta into longitudinal and transverse parts in
the high-energy limit is
\begin{eqnarray}
&&q_1=x_1p_1+q_{1,t},\qquad q_2=x_2p_2+q_{2,t},\qquad 0<x_{1,2}<1, \\
&&q_0=x'_1p_1+x'_2p_2+q_{0,t},\quad x'_1\sim x'_2\ll x_{1,2},\quad
q_{0,1,2}^2\simeq q_{0/1/2,t}^2. \nonumber \label{dec}
\end{eqnarray}
Making use of energy-momentum conservation laws
\begin{eqnarray}
q_1=p_1-p'_1-q_0,\qquad q_2=p_2-p'_2+q_0,\qquad q_1+q_2=k_1+k_2
\label{CL}
\end{eqnarray}
we write
\begin{eqnarray}\label{sx1x2}
s\,x_1x_2=M_{q\bar{q}}^2+|{\bf P}_{t}|^2\equiv
M_{q\bar{q},\perp}^2,\qquad M_{q\bar{q}}^2=(k_1+k_2)^2  \; ,
\end{eqnarray}
where $M_{q\bar{q}}$ is the invariant mass of the $q\bar{q}$ pair,
and ${\bf P}_{t}$ is its transverse 3-momentum.

\subsection{The amplitude for $p p \to p p Q \bar Q$}

Let us concentrate on the simplest case of production of $q\bar{q}$
pair in the color singlet state. Color octet state would demand an
emission of an extra gluon \cite{KMR-g} which considerably
complicates the calculations, and we postpone such an analysis for
future studies.

In analogy to the Kaidalov-Khoze-Martin-Ryskin approach (KKMR)
\cite{KMR_Higgs,KKMR,KKMR-spin} for Higgs boson production, we write
the amplitude of the exclusive diffractive $q\bar{q}$ pair
production $pp\to p(q\bar{q})p$ in the color singlet state as
\begin{eqnarray}
{\cal
M}_{\lambda_q\lambda_{\bar{q}}}^{p p \to p p q \bar q}(p'_1,p'_2,k_1,k_2) &=&
s\cdot\pi^2\frac12\frac{\delta_{c_1c_2}}{N_c^2-1}\,
\Im\int d^2
q_{0,t} \; V_{\lambda_q\lambda_{\bar{q}}}^{c_1c_2}(q_1, q_2, k_1, k_2) \nonumber \\
&&\frac{f^{\mathrm{off}}_{g,1}(x_1,x_1',q_{0,t}^2,
q_{1,t}^2,t_1)f^{\mathrm{off}}_{g,2}(x_2,x_2',q_{0,t}^2,q_{2,t}^2,t_2)}
{q_{0,t}^2\,q_{1,t}^2\, q_{2,t}^2} \; ,
\label{amplitude}
\end{eqnarray}
where $\lambda_q,\,\lambda_{\bar{q}}$ are helicities of heavy $q$
and $\bar{q}$, respectively. Above $f_1^{\mathrm{off}}$ and
$f_2^{\mathrm{off}}$ are the off-diagonal unintegrated gluon
distributions in nucleon 1 and 2, respectively. They will be
discussed in a separate subsection below.

The longitudinal momentum fractions of active gluons
are calculated based on kinematical variables of outgoing quark
and antiquark
\begin{eqnarray}
x_1 &=& \frac{m_{3,t}}{\sqrt{s}} \exp(+y_3)
     +  \frac{m_{4,t}}{\sqrt{s}} \exp(+y_4) \; , \nonumber \\
x_2 &=& \frac{m_{3,t}}{\sqrt{s}} \exp(-y_3)
     +  \frac{m_{4,t}}{\sqrt{s}} \exp(-y_4) \; ,
\label{x_1_x_2}
\end{eqnarray}
where $m_{3,t}$ and $m_{4,t}$ are transverse masses of the quark and
antiquark, respectively, and $y_3$ and $y_4$ are corresponding
rapidities.

The bare amplitude above is subjected to absorption corrections
which, in general, depend on collision energy and on the spin-parity
of the produced central system \cite{KMR_chic}. We shall discuss
this issue shortly when presenting our results.

\subsection{$g g \to Q \bar Q$ vertex}

Let us consider the subprocess amplitude for the $q\bar{q}$ pair production
via off-shell gluon-gluon fusion. The vertex factor
$V_{\lambda_q\lambda_{\bar{q}}}^{c_1c_2}=
 V_{\lambda_q\lambda_{\bar{q}}}^{c_1c_2}(q_1,q_2,k_1,k_2)$
in expression (\ref{amplitude}) is the production amplitude
of a pair of massive quark $q$ and antiquark $\bar{q}$ with
helicities $\lambda_q$, $\lambda_{\bar{q}}$ and
momenta $k_1$, $k_2$, respectively.
Within the QMRK approach \cite{FL}, the
color singlet $q\bar{q}$ pair production amplitude can be written
as
\begin{eqnarray}\label{qqamp}
&&V_{\lambda_q\lambda_{\bar{q}}}^{c_1c_2}(q_1,q_2,k_1,k_2)\equiv
n^+_{\mu}n^-_{\nu}V_{\lambda_q\lambda_{\bar{q}}}^{c_1c_2,\,\mu\nu}(q_1,q_2,k_1,k_2), \\
&&V_{\lambda_q\lambda_{\bar{q}}}^{c_1c_2,\,\mu\nu}(q_1,q_2,k_1,k_2)
=-g^2\sum_{i,k}\left\langle 3i,\bar{3}k|1\right\rangle\times
\nonumber \\
&&\bar{u}_{\lambda_q}(k_1)
(t^{c_1}_{ij}t^{c_2}_{jk}b^{\mu\nu}(q_1,q_2,k_1,k_2)-
t^{c_2}_{kj}t^{c_1}_{ji}\bar{b}^{\mu\nu}(q_1,q_2,k_1,k_2))v_{\lambda_{\bar{q}}}(k_2),
\nonumber
\end{eqnarray}
where $t^c$ are the color group generators in the fundamental
representation, $u(k_1)$ and $v(k_2)$ are on-shell quark and
antiquark spinors, respectively, $b,\,\bar{b}$ are
vertices (\ref{bb}) arising from the Feynman rules
:
\begin{eqnarray} \label{bb}
b^{\mu\nu}(q_1,q_2,k_1,k_2)=\gamma^{\nu}\frac{\hat{q}_{1}-\hat{k}_{1}-m}{(q_1-k_1)^2-m^2}
\gamma^{\mu}
\; , \\
\bar{b}^{\mu\nu}(q_1,q_2,k_1,k_2)=\gamma^{\mu}\frac{\hat{q}_{1}-\hat{k}_{2}+m}{(q_1-k_2)^2-m^2}
\gamma^{\nu}
\; . \nonumber
\end{eqnarray}
%

The SU(3) Clebsch-Gordan coefficient $\left\langle
3i,\bar{3}k|1\right\rangle=\delta^{ik}/\sqrt{N_c}$ in
Eq.~(\ref{qqamp}) projects out the color quantum numbers of the
$q\bar{q}$ pair onto the color singlet state. Factor $1/\sqrt{N_c}$
provides the averaging of the matrix element squared over
intermediate color states of quarks.

The tensorial part of the amplitude is therefore:
\begin{eqnarray}\nonumber
&&{}V_{\lambda_q\lambda_{\bar{q}}}^{\mu\nu}(q_1, q_2, k_1, k_2)
= g_s^2 \,\bar{u}_{\lambda_q}(k_1)
\biggl(\gamma^{\nu}\frac{\hat{q}_{1}-\hat{k}_{1}-m}
{(q_1-k_1)^2-m^2}\gamma^{\mu}-\gamma^{\mu}\frac{\hat{q}_{1}-
\hat{k}_{2}+m}{(q_1-k_2)^2-m^2}\gamma^{\nu}\biggr)v_{\lambda_{\bar{q}}}(k_2).\\
\label{vector_tensor}
\end{eqnarray}

Taking into account definitions (\ref{lcbasis}) and momentum
conservation (\ref{dec}) and using the gauge invariance properties
we get the following projection to the light cone vectors
(so called ``Gribov's trick'')
\begin{eqnarray}
V_{\lambda_q\lambda_{\bar{q}}}^{c_1c_2}(q_1, q_2, k_1, k_2)
&=& n^+_{\mu}n^-_{\nu}
V_{\lambda_q\lambda_{\bar{q}},\,\mu\nu}^{c_1c_2}(q_1, q_2, k_1,k_2) \nonumber \\
&=&
\frac{4}{s}\frac{q^{\nu}_1-q^{\nu}_{1,t}}{x_1}\frac{q^{\mu}_2-q^{\mu}_{2,t}}{x_2}
V^{c_1c_2}_{\lambda_q\lambda_{\bar{q}},\,\mu\nu}(q_1,q_2,k_1,k_2)
\nonumber \\
&=& \frac{4}{s}\frac{q^{\nu}_{1,t}}{x_1}\frac{q^{\mu}_{2,t}}{x_2}
V^{c_1c_2}_{\lambda_q\lambda_{\bar{q}},\,\mu\nu}(q_1,q_2,k_1,k_2).
\label{trick}
\end{eqnarray}

Using (\ref{vector_tensor}) and (\ref{trick}) we can write
\begin{eqnarray}
V_{\lambda_q\lambda_{\bar{q}}}^{c_1c_2}(q_1, q_2, k_1, k_2) =
\frac{4}{s x_1 x_2} g_s^2 \,\bar{u}_{\lambda_q}(k_1)
\biggl(\hat{q}_{1t} \frac{\hat{q}_{1}-\hat{k}_{1}-m}
{(q_1-k_1)^2-m^2}\hat{q}_{2t}
-\hat{q}_{2t} \frac{\hat{q}_{1}-
\hat{k}_{2}+m}{(q_1-k_2)^2-m^2}\hat{q}_{1t} \biggr)
v_{\lambda_{\bar{q}}}(k_2) \nonumber . \\
\label{vertex_function_formula}
\end{eqnarray}
The coupling constants $g_s^2 \to g_s(\mu_{r,1}^2)
g_s(\mu_{r,2}^2)$. In the present calculation we take the
renormalization scale to be $\mu_{r,1}^2=\mu_{r,2}^2=M_{q \bar
q}^2/4$. The matrix element (\ref{vertex_function_formula}) is then
calculated numerically. Inserting it to Eq.~(\ref{amplitude}) we can
calculate numerically the whole amplitude for the $p p \to p p Q
\bar Q$ process.

\subsection{Off-diagonal unintegrated gluon distributions}

In the KMR approach the off-diagonal parton distributions
are calculated as
\begin{eqnarray}
f_1^{\mathrm{KMR}}(x_1,Q_{1,t}^2,\mu^2,t_1) &=& R_g
\frac{d[g(x_1,k_t^2)S_{1/2}(k_{t}^2,\mu^2)]}{d \log k_t^2} |_{k_t^2
= Q_{1t}^2} \;
F(t_1) \;  \nonumber \\
&\approx& R_g \frac{dg(x_1,k_t^2)}{d \log k_t^2} |_{k_t^2 =
Q_{1,t}^2} \;
S_{1/2}(Q_{1,t}^2,\mu^2) \; F(t_1) \; , \nonumber \\
f_2^{\mathrm{KMR}}(x_2,Q_{2,t}^2,\mu^2,t_2) &=& R_g
\frac{d[g(x_2,k_t^2)S_{1/2}(k_{t}^2,\mu^2)]}{d \log k_t^2} |_{k_t^2
= Q_{2t}^2} \;
F(t_2) \;  \nonumber \\
&\approx& R_g \frac{dg(x_2,k_t^2)}{d \log k_t^2} |_{k_t^2 =
Q_{2,t}^2} \;
S_{1/2}(Q_{2,t}^2,\mu^2) \; F(t_2) \; , \nonumber \\
\label{KMR_off-diagonal-UGDFs}
\end{eqnarray}
where $S_{1/2}(q_t^2, \mu^2)$ is a Sudakov-like form factor relevant
for the case under consideration \cite{MR}. The last approximate(!)
equalities come from the fact that in the region under consideration
the Sudakov-like form factors are somewhat slower functions of
transverse momenta than the collinear gluon distributions. While
reasonable for an estimate of gluon distribution it may be not
sufficient for precise calculation of the cross section. It is
reasonable to take a running (factorization) scale as: $\mu_1^2 =
\mu_2^2 = M_{q \bar q}^2/4$. We shall call the formulae
(\ref{KMR_off-diagonal-UGDFs}) as the DDT-like formulae \cite{DDT},
for brevity.

The factor $R_g$ here cannot be calculated from first principles
in the most general case of off-diagonal UGDFs.
It can be estimated in the case of off-diagonal collinear PDFs
when $x' \ll x$ and $x g = x^{-\lambda}(1-x)^n$ \cite{Shuvaev_PDF}.
Then
\begin{equation}
R_g = \frac{2^{2\lambda+3}}{\sqrt{\pi}}
\frac{\Gamma(\lambda+5/2)}{\Gamma(\lambda+4)} \; .
\label{R_g}
\end{equation}
In a more realistic case of DGLAP GDF $\lambda = \lambda(x,\mu^2)$.
Typically $R_g \sim$ 1.3 -- 1.4 at Tevatron energy. A more general
case of unintegrated off-diagonal distributions was discussed in
Ref.~\cite{MR}, but we will not touch them here.

The off-diagonal form factors are parametrized here as:
\begin{equation}
F(t) = \exp \left( B_{\mathrm{off}} t \right) \; .
\end{equation}
In practical calculations in this letter we take $B_{\mathrm{off}}$
= 2 GeV$^{-2}$.

In the original KMR approach the following prescription for the
effective transverse momentum is taken:
\begin{eqnarray}
Q_{1,t}^2 &=& \min\left( q_{0,t}^2,q_{1,t}^2 \right) \; , \nonumber \\
Q_{2,t}^2 &=& \min\left( q_{0,t}^2,q_{2,t}^2 \right) \; .
\label{effective_transverse_momenta}
\end{eqnarray}
Other prescriptions are also possible \cite{PST_chic01}.

In evaluating $f_1$ and $f_2$ needed for calculating the amplitude
(\ref{amplitude}) we use the GRV collinear distributions
\cite{GRV95}.

\begin{figure}[h!]
\centerline{\epsfig{file=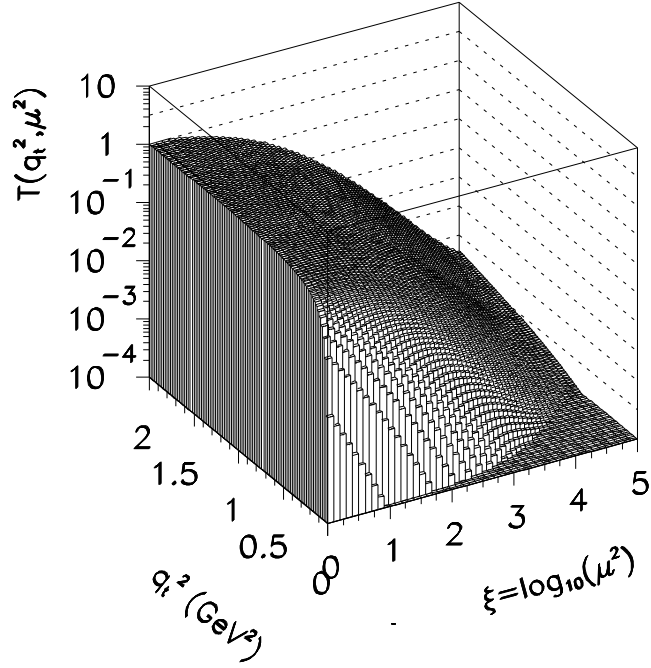,width=10cm}}
\caption{\small The Sudakov form factor as a function of transverse
momentum squared $q_t^2$ and a log$_{10}$ of the scale parameter
$\mu^2$.}
\label{fig:sudakov_map}
\end{figure}

It was proposed \cite{MR} to express the $S_{1/2}$ form factors in
Eq.~(\ref{KMR_off-diagonal-UGDFs}) through the standard Sudakov form
factors as:
\begin{equation}
S_{1/2}(q_t^2,\mu^2) = \sqrt{T_g(q_t^2,\mu^2)}  \; .
\label{S_{1/2}}
\end{equation}
The Sudakov form factor, a two-dimensional function
$T_g(q_t^2,\mu^2)$ as a function of transverse momentum squared
$q_t^2$ and a log$_{10}$ of the scale parameter $\mu^2$, is shown in
Fig.~\ref{fig:sudakov_map}.

A strong dependence on $\mu^2$ ($\log_{10}(\mu^2)$ in the figure) is
clearly visible. This dependence leads to a huge perturbative
damping of the off-diagonal UGDFs (and, as a consequence, of the
amplitude and the cross section) in the case when objects $X$ with
sizeable masses ($M_X$) are produced, i.e. when $\mu^2 \sim M_X^2$.

\subsection{The $pp \to pp Q \bar Q$ cross section}

The cross section is obtained by assuming a general
$2 \to 4$ reaction:
\begin{equation}
d \sigma = \frac{1}{2s} |{\cal M}_{2 \to 4}|^2 (2 \pi)^4
\delta^4(p_a + p_b - p_1 - p_2 - p_3 - p_4) \frac{d^3 p_1}{(2 \pi)^3
2 E_1} \frac{d^3 p_2}{(2 \pi)^3 2 E_2} \frac{d^3 p_3}{(2 \pi)^3 2
E_3} \frac{d^3 p_4}{(2 \pi)^3 2 E_4} \; . \label{phase_space}
\end{equation}
The details how to conveniently reduce the number of
kinematical integration variables are given elsewhere \cite{LS09}.

\section{Results}

As for the exclusive production of $\chi_c$ mesons
\cite{KMR_chic,PST_chic2}, in the case when the KMR UGDFs are used
an extra cut-off on transverse momenta is applied i.e. the formula
(\ref{amplitude}) is used if
\begin{eqnarray}
Q_{1,t}^2 &=& \min(q_{0,t}^2,q_{1,t}^2) > Q_{t,cut}^2 \; ,
\nonumber \\
Q_{2,t}^2 &=& \min(q_{0,t}^2,q_{2,t}^2) > Q_{t,cut}^2 \; .
\end{eqnarray}
Otherwise the cross section is set to zero.

Let us proceed now with the presentation of differential
distributions of charm quarks produced in the EDD mechanism. In our
calculation here we fix the scale of the Sudakov form factor to be
$\mu = M_{c \bar c}/2$. Such a choice of the scale leads to a strong
damping of the situations with large rapidity gaps between $c$ and
$\bar c$ \footnote{Large rapidity gap means automatically large
invariant mass $M_{c \bar c}$ of the $c \bar c$ system.}.

\begin{figure}[h!]
\includegraphics[width=0.6\textwidth]{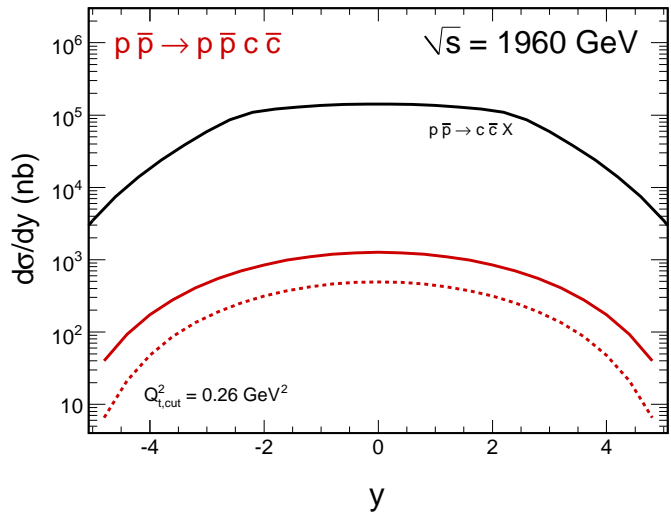}
\caption{\small Rapidity distribution of $c$ or $\bar c$. The upper
curve is for inclusive production in the $k_t$-factorization
approach with the Kwieci\'nski UGDF and $\mu^2 = 4 m_c^2$, while the
two lower lines are for the EDD mechanism for the KMR UGDF with
leading-order collinear gluon distribution \cite{GRV95}. The solid
line is calculated from the exact DDT-like formula (see
Eq.~(\ref{KMR_off-diagonal-UGDFs})) and the dashed line for the
simplified formula (when only derivative of collinear GDF is taken).
An extra cut on the momenta in the loop $Q_{t,cut}^2$ = 0.26 GeV$^2$
was imposed. Absorption effects were included approximately by
multiplying the cross section by the gap survival factor $S$ = 0.1.
} \label{fig:dsig_dy}
\end{figure}

In Fig.~\ref{fig:dsig_dy} we show distribution in rapidity. The
results obtained with the KMR method are shown together with
inclusive gluon-gluon contribution calculated as in
Ref.~\cite{LS06}.
The effect of absorption leads to a damping
of the cross section by an energy-dependent factor. For
the Tevatron this factor is about 0.1. If the extra factor
is taken into account the EDD contribution is of the order
of 1\% of the dominant gluon-gluon fusion contribution.

The corresponding rapidity-integrated cross sections are: 6.6 $\mu$b
for exact DDT formula, 2.4 $\mu$b for simplified DDT formula (see
Eq.~(\ref{KMR_off-diagonal-UGDFs})). For comparison the inclusive
cross section (gluon-gluon component only) is 807 $\mu$b.

\begin{figure}[h!]
\includegraphics[width=0.6\textwidth]{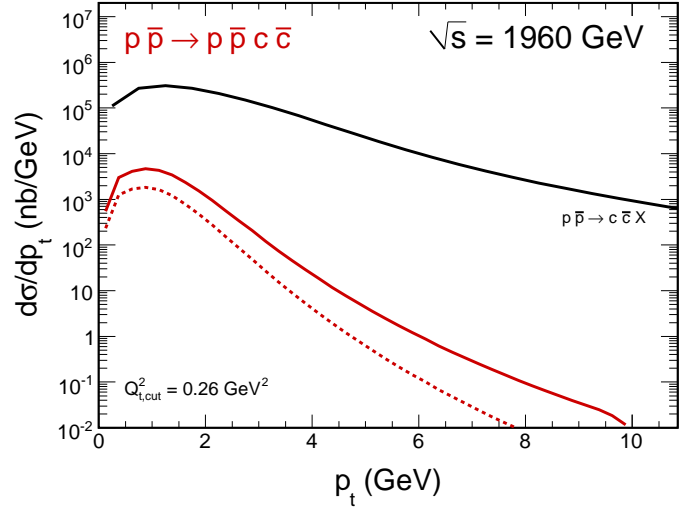}
\caption{\small Transverse momentum distribution of $c$ or $\bar c$.
The other details are the same as in Fig.~\ref{fig:dsig_dy}. }
\label{fig:dsig_dpt}
\end{figure}

In Fig.~\ref{fig:dsig_dpt} we show the differential cross section in
transverse momentum of the charm quark. Compared to the inclusive
case, the exclusive contribution falls significantly faster with
transverse momentum than in the inclusive case.

\begin{figure}[h!]
\includegraphics[width=0.6\textwidth]{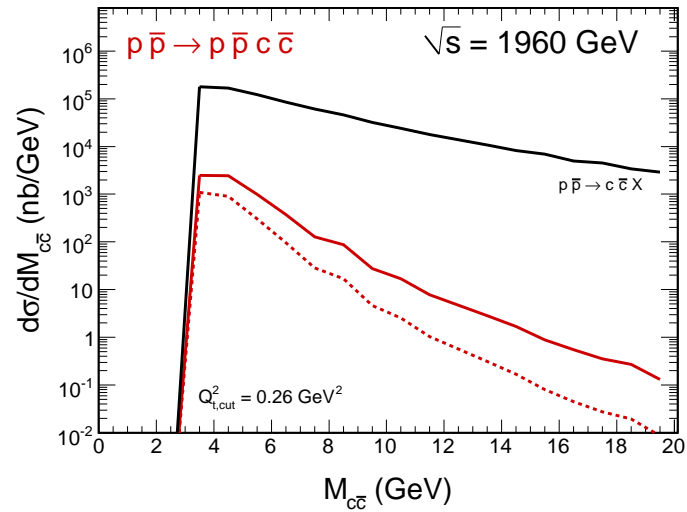}
\caption{\small Invariant mass distribution of the $c \bar c$ pair.
The other details are the same as in Fig.~\ref{fig:dsig_dy}. }
\label{fig:dsig_dm34}
\end{figure}

In Fig.~\ref{fig:dsig_dm34} we show the distribution in the
invariant mass of $c$ and $\bar c$. The fluctuations visible in the
figure are due to the fact that the integration is not directly in
$M_{c \bar c}$, but in other variables, and the number of
integration points is rather restricted. Compared to the inclusive
case the invariant mass distribution for the EDD component is
significantly steeper. This is due to the Sudakov-like form factor
which, according to the procedure described above, damps the cross
section for large invariant masses $M_X$.

\begin{figure}[h!]
\includegraphics[width=0.6\textwidth]{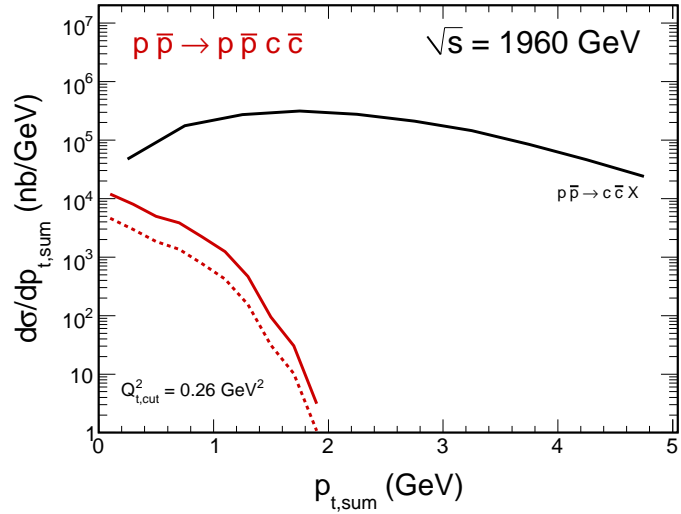}
\caption{\small Distribution in the transverse momentum of the $c
\bar c$ pair. The other details are the same as in
Fig.~\ref{fig:dsig_dy}. } \label{fig:dsig_dptsum}
\end{figure}

As in the inclusive case within the $k_t$-factorisation approach
\cite{LS06} the $c \bar c$ pair possesses the transverse momentum
different from zero. The corresponding distribution is shown in
Fig.~\ref{fig:dsig_dptsum}. The distribution for the exclusive case
(the two lower lines) is much narrower compared to the inclusive
case (the upper line).

\section{Conclusions}

In the present letter we have evaluated, for the first time in the literature,
the contribution of exclusive-double diffractive production
of open charm.
We have found a sizeable cross sections of
the order of 1 \% of the standard inclusive
gluon-gluon fusion contribution at the Tevatron energy.
The details depend, however, on the UGDFs used in
the evaluation of the cross section.
The most reliable estimate is obtained with the KMR
off-diagonal UGDFs. These distributions were verified
recently in the production of $\chi_c$ quarkonia
\cite{KMR_chic,PST_chic2}, i.e. for the kinematics similar to
the present one.

It would be therefore very valuable to measure the diffractive
mechanism discussed in the present paper. How to identify the EDD
contribution? The events corresponding to the EDD contribution are
expected to be related to a rather small multiplicity of particles
(mainly pions or kaons) associated with $c$ or $\bar c$, or more
precisely charmed mesons which are formed in the process of
hadronization of charm quarks into charmed mesons. One method would
be therefore to measure $D$ mesons with a trigger on small pion
multiplicity. Another method would be to measure $D$ mesons in
association with rapidity gaps with respect to the outgoing
protons/antiprotons. This is partially possible at the Tevatron and
will be accessible at the LHC as well.

\vspace{1cm}

{\bf Acknowledgments}

We are indebted to Ch. Royon for an interesting discussion. Useful
discussions with Gunnar Ingelman and Oleg Teryaev are gratefully
acknowledged. This study was partially supported by the Polish grant
of MNiSW N N202 249235.



\begin{thebibliography}{100}

\bibitem{NLO_ccbar}
P. Nason, S. Dawson and R.K. Ellis, Nucl. Phys. {\bf B303} (1988) 607;\\
G. Altarelli, M. Diemoz, G. Martinelli and P. Nason,
Nucl. Phys. {\bf B308} (1988) 724;\\
P. Nason, S. Dawson and R.K. Ellis, Nucl. Phys. {\bf B327} (1989) 49;\\
W. Beenakker, H. Kuijf, W.L. Van Neerven and J. Smith, Phys. Rev. {\bf D40}
(1989) 54.

\bibitem{kt-factorisation_ccbar}
S. Catani, M. Ciafaloni and F. Hautmann, Nucl. Phys. {\bf 366} (1991)
135; \\
J.C. Collins and R.K. Ellis, Nucl. Phys. {\bf B360} (1991) 3;\\
R.D. Ball and R.K. Ellis, J.H.E.P. {\bf 0105} (2001) 053.

\bibitem{BS00}
S.P. Baranov and M. Smizanska, Phys. Rev. {\bf D62} (2000) 014012.

\bibitem{HKSST-qq}
  P.~Hagler, R.~Kirschner, A.~Schafer, L.~Szymanowski and O.~Teryaev,
  Phys.\ Rev.\  D {\bf 62}, 071502 (2000)
  [arXiv:hep-ph/0002077].

\bibitem{LS06}
M. {\L}uszczak and A. Szczurek,
Phys. Rev. {\bf D73} (2006) 054028.

\bibitem{KMR_Higgs_bbbar_background}
  A.~De.~Roeck et al., Eur. Phys. J. {\bf C25} (2002) 391;\\
  S.~Heinemeyer et al., Eur. Phys. J. {\bf C53} (2008) 231;\\
  A.~G.~Shuvaev, V.~A.~Khoze, A.~D.~Martin and M.~G.~Ryskin,
  Eur.\ Phys.\ J.\  C {\bf 56}, 467 (2008)
  [arXiv:0806.1447 [hep-ph]].

\bibitem{KMR_Higgs}
V.A. Khoze, A.D. Martin and M.G. Ryskin, Phys. Lett. B {\bf 401},
330 (1997);\\
V.A. Khoze, A.D. Martin and M.G. Ryskin, Eur. Phys. J. C {\bf 23},
311 (2002).

\bibitem{KKMR}
A.B. Kaidalov, V.A. Khoze, A.D. Martin and M.G. Ryskin, Eur. Phys.
J. C {\bf 33}, 261 (2004).

\bibitem{KKMR-spin}
A.B. Kaidalov, V.A. Khoze, A.D. Martin and M.G. Ryskin, Eur.\ Phys.\
J.\ C {\bf 31}, 387 (2003) [arXiv:hep-ph/0307064].

\bibitem{PST_chic01}
  R.~S.~Pasechnik, A.~Szczurek and O.~V.~Teryaev,
  Phys.\ Rev.\  D {\bf 78}, 014007 (2008)
  [arXiv:0709.0857 [hep-ph]];\\
  R.~S.~Pasechnik, A.~Szczurek and O.~V.~Teryaev,
  Phys.\ Lett.\  B {\bf 680}, 62 (2009)
  [arXiv:0901.4187 [hep-ph]].

\bibitem{KMR_chic}
L.A. Harland-Lang, V.A. Khoze, M.G. Ryskin and W.J. Stirling,
arXiv:0909.4748.

\bibitem{PST_chic2}
R. Pasechnik, A. Szczurek and O.V. Teryaev, arXiv:0912.4251[hep-ph].

\bibitem{KMR-g}
  V.~A.~Khoze, M.~G.~Ryskin and A.~D.~Martin,
  Eur.\ Phys.\ J.\  C {\bf 64}, 361 (2009)
  [arXiv:0907.0966 [hep-ph]].

\bibitem{FL}
  V.~S.~Fadin and L.~N.~Lipatov,
  Nucl.\ Phys.\  B {\bf 477}, 767 (1996)
  [arXiv:hep-ph/9602287].

\bibitem{HKSST-charm}
  P.~Hagler, R.~Kirschner, A.~Schafer, L.~Szymanowski and O.~V.~Teryaev,
  Phys.\ Rev.\ Lett.\  {\bf 86}, 1446 (2001)
  [arXiv:hep-ph/0004263].

\bibitem{GRV95}
M. Gl\"uck, E. Reya and A. Vogt, Z. Phys. {\bf C67} (1995) 433.

\bibitem{MR}
  A.~D.~Martin and M.~G.~Ryskin,
  Phys.\ Rev.\  D {\bf 64}, 094017 (2001)
  [arXiv:hep-ph/0107149].

\bibitem{DDT}
Yu.L. Dokshitzer, D.I. Dyakonov and S.I. Troyan,
Phys. Rep. {\bf 58} (1980) 269.

\bibitem{LS09}
P. Lebiedowicz and A. Szczurek, arXiv:0912.0190[hep-ph].

\bibitem{Shuvaev_PDF}
A.G. Shuvaev et al., Phys. Rev. {\bf D60} (1999) 014015.


\end{thebibliography}
\end{document}